\documentstyle[11pt]{article}
\newcommand{\al}{\alpha}

\newcommand{\del}{\delta}
\newcommand{\ep}{\epsilon}

\newcommand{\th}{\theta}

\newcommand{\la}{\lambda}

\newcommand{\si}{\sigma}

\newcommand{\om}{\omega}

\newcommand{\De}{\Delta}

\newcommand{\La}{\Lambda}

\newcommand{\Om}{\Omega}

\newcommand{\be}{\begin{eqnarray}}
\newcommand{\ee}{\end{eqnarray}}
\newcommand{\jt}{\tilde{J}}

\newcommand{\lra}{\longrightarrow}
\newcommand{\pr}{\partial}
\newcommand{\ti}{\tilde}

\newcommand{\ran}{\rangle}
\newcommand{\bz}{\bar{z}}
\newcommand{\bJ}{\bar{J}}

\newcommand{\ind}{\indent}
\newcommand{\np}{\newpage}
\newcommand{\hs}{\hspace}
\newcommand{\vs}{\vspace}

\newcommand{\nn}{\nonumber}
\newcommand{\lef}{\left}
\newcommand{\rig}{\right}

\begin{document}

\batchmode

\thispagestyle{empty}

\begin{flushright}
G\"oteborg ITP 94-16 \\
June 1994 \\
hep-th 9407082 \\
\end{flushright}
\vs{5mm}

\begin{center}

{\huge{BRST invariant branching functions of G/H coset models.}} \\
\vspace{10 mm}
{\Large{Henric Rhedin}\footnote{hr@fy.chalmers.se}} \\
\vspace{2 mm}

Institute of Theoretical Physics \\
G\"oteborg University \\
and \\
Chalmers University of Technology \\

\vs{15mm}

{\bf{Abstract}} \\

\end{center}

We compute branching functions of $G/H$ coset models using a BRST
invariant branching function formulae, i.e.
a branching function that respects a BRST invariance of the model.
This ensures that only the coset degrees of freedom will
propagate. We consider $G/H$ for rank$(G/H)=0$ models which includes
the Kazama-Suzuki
construction, and $G_{k_1}\times G_{k_2}/G_{k_1+k_2}$
models.
Our calculations here confirm in part previous results for
those models which have been obtained under an assumption
in a free field approach.
We also consider $G_{k_1}\times H_{k_2}/H_{k_1+k_2}$,
where $H$ is a subgroup of $G$, and
$\prod_{a=1}^mG_{k_a}/G_{\sum_{a=1}^nk_a}$, whose branching functions,
to our knowledge, has not been calculated before.

\np

\setcounter{page}{1}

\section{Introduction}

Two dimensional conformal field theories (CFT's) has been
thoroughly investigated due to its applications in
both string theories and in the study of critical
phenomena. The internal degrees of freedom for a string theory
are described by a unitary CFT. \\
It is generally believed that all CFT's are described by the
Wess-Zumino-Novikov-Witten (WZNW) model
\cite{Witten84},\cite{Gepner-Witten86},\cite{Knizhnik-Zamolodchikov89},
or gauged versions of it
\cite{Gawedzki-Kupiainen89},\cite{Karabali-Park-Schnitzer-Yang89}.
The action of
the WZNW model is constructed from fields that take values
in some group $G$. From the stress-energy tensor associated with
the group $G$ WZNW model, Goddard-Kent-Olive (GKO)
\cite{Goddard-Kent-Olive85}
constructed more general stress-energy tensors
using a so-called coset construction. On the level of states
this essentially means that the states of the state-space
should all be primary with respect to the currents of
some subgroup $H$ of $G$. \\
\indent The stress-energy tensor of the GKO construction may also be
found from the a BRST analysis of the gauged WZNW model
\cite{Karabali-Schnitzer84}. One essential ingredient in this construction
is the introduction of an auxiliary sector $\ti{H}$.
In \cite{Hwang-Rhedin93} the equivalence of the
two approaches was shown for integrable representations of
the group $G$. \\
\indent We will in this paper concentrate on calculating
branching functions of various coset models using the BRST approach.
In \cite{Hwang-Rhedin93} a branching function formula that respects the BRST
invariance was given. The BRST analysis shows \cite{Hwang-Rhedin93},
that the remaining degrees of freedom are contained in the coset
sector, and having a BRST invariant branching function thus
ensures us that only the coset degrees of freedom are propagating.
The BRST invariant branching function is decomposable into three parts;
a character for the entire group $G$, a character for the
auxiliary sector $\ti{H}$ and one for the Faddev-Popov ghosts.
Since the BRST invariant
branching function contains the character of the full group $G$, it respects
the invariances of this group. This means that we have a
branching function which is independent of the decomposition of $G$ in
$H$, and this will make it possible to compute the branching functions
explicitly in a straightforward manner.
This is not the case in for instance \cite{Goddard-Kent-Olive86}, where the
Virasoro minimal model branching function,
$SU(2)_k\times SU(2)_1/SU(2)_{k+1}$, is calculated.   \\
\indent Branching functions or coset characters\footnote{
We will in what follows, somewhat inconsitently, use the notation
character instead of branching function. Strictly speaking those
are not the same. We are most grateful to Prof. B. Schellekens
for clearifying this point to us.} have previously been
calculated for a number of special cases,
\cite{Goddard-Kent-Olive86},\cite{Gepner-Qiu87},\cite{Distler-Qiu90},
\cite{Ravanini88},\cite{Ravanini-Yang87},\cite{Kastor-Martinec-Qiu88},
\cite{Bagger-Nemeschansky-Yankielowicz88},
in general confined to various coset constructions of the groups
$SU(2)$ and $U(1)$, as well as for general constructions,
\cite{Bouwknegt-McCarthy-Pilch91},\cite{Christe-Ravanini89},
\cite{Huito-Nemeschansky-Yankielowicz90}.
We have for example characters for the Kazama-Suzuki
\cite{Kazama-Suzuki89:2} models,
calculated in \cite{Bouwknegt-McCarthy-Pilch91}. The
constructions of reference \cite{Bouwknegt-McCarthy-Pilch91}
are, however, computed using an assumption.
Since we are provided with a way to compute characters
without assumptions, we can make interesting comparisons.
References \cite{Christe-Ravanini89},\cite{Huito-Nemeschansky-Yankielowicz90}
give exact results but
consider only one general model ( ref. \cite{Christe-Ravanini89} ) or
calculate only a few examples ( ref.
\cite{Huito-Nemeschansky-Yankielowicz90} ).
Also in
\cite{Christe-Ravanini89} the branching functions are not given in an
explicit form.
We will in fact, to some extent, show coincidence with those cases. \\
\ind We will also give coset characters for models that have not
previously been considered in literature, at least not in
the general case.

\section{Preliminaries}

We start by a short review of the background and refer to
\cite{Hwang-Rhedin93}, and references therein for more details.
Consider a WZNM model on a Riemann surface ${\cal M}$ with
fields g taking values in a compact Lie group $G$.  The
action is \cite{Witten84},\cite{Gepner-Witten86} and
\cite{Knizhnik-Zamolodchikov89}
\be
S_k=\frac{k}{16\pi}\int_{\cal M}d\si d\tau Tr(\pr_{\mu}{\rm g\pr^{\mu}
g^{-1}})+
\frac{k}{24\pi}\int_B d^3y\ep^{abc}Tr({\rm g}^{-1}\pr_a{\rm g}
{\rm g}^{-1}
\pr_b{\rm g}{\rm g}^{-1}\pr_c{\rm g}), \label{wznwaction}
\ee
where ${\cal M}$ is the boundary of $B$ on which g is
supposed to be well defined.  $k$ is referred to as the level
of the WZNM model.  In general, every simple part of $G$ may
have different levels but we take for the moment $G$ simple. \\
The action is invariant under the transformations
\be
{\rm g}(z,\bz)\lra\bar{\Om}^{-1}(\bz){\rm g}(z,\bz)\Om(z). \label{wznwinv}
\ee
$\Om$ and $\bar{\Om}$ are arbitrary analytical group-valued
matrices.  This symmetry gives rise to an infinite number
of conserved currents which are found to be of the affine
Lie type.  That is, they transform as
\be
\del_{\om}J(z)=[J(z),\om(z)]-\frac{k}{2}\pr_z\om(z) \nn \\
\del_{\bar{\om}}\bJ(\bz)=[\bJ(\bz),\bar{\om}(\bz)]-
\frac{k}{2}\pr_{\bz}\bar{\om}(\bz) \label{wznwtrans}
\ee
under the infinitesimal version of eq.(\ref{wznwinv}).  Making a
Laurent expansion this gives the well known affine Lie
algebra, $\hat{g}$, of level $k$,
\be [J_m^A,J_n^B]=if^{AB}_{\ \ C}J^c_{m+n}+
\frac{k}{2}mg^{AB}\del_{m+n,0}. \label{kacmoody}
\ee
Likewise for $\bar{J}$.  $f^{AB}_{\ \ C}$ are structure
constants of the Lie algebra $g$ of $G$.  $g^{AB}$ is a
non-degenerate metric on $G$.  The Sugawara stress-energy
tensor splits into holomorphic and anti-holomorphic parts,
and in the former case it becomes
\be
T(z)&=&\frac{1}{k+c_G}:J^A(z)J_A(z): \nn \\
&=&\frac{1}{k+c_G}\sum_{\stackrel{\scriptstyle{m\in{\cal Z}}}
{n\in{\cal Z}}
}:J^A_{n-m}J_{A,m}:z^{-n-2} \label{stress},
\ee
where $c_G$ is the quadratic Casimir of the adjoint
representation of $g$.  The corresponding central
charge becomes
\be
c=\frac{kd_G}{k+c_G}, \label{confcha}
\ee
where $d_G$ is the dimension of $G$. \\

In reference \cite{Goddard-Kent-Olive85} it was described how to
enlarge the range of central charges of models with
stress-energy tensors given in the form of eq.(\ref{stress}).
Taking a subgroup $H$ of $G$
one may note that the difference of the Virasoro
generators of $G$ and $H$ also obeys a Virasoro algebra of
central charge
\be
c^{tot}=\frac{kd_G}{k+c_G}-\frac{kd_H}{k+c_H}
\ee
for a simple subgroup $H$.
Since the currents of the subgroup $H$ commutes with the
stress-energy tensor it is natural to impose the
constraint\footnote{We have adapted the convention
that uppercase letters take values in $g$ while
lowercase take values in $h$.}
\be
J^a_n|{\rm phys}\rangle=0, \hs{10mm} n>0 \hs{5mm} {\rm or}\ \ n=0\ \
{\rm and} \ \ a\in\De^+_h \label{coset}
\ee
on the physical states.  This we will refer to as the GKO
coset condition. \\

In order to gauge the WZNW model we use an anomaly-free
subgroup $H$ of $G$.  We introduce gauge fields which
transforms in the adjoint representation of $H$.
The gauged action in light-cone coordinates looks like
\be
S_{k}({\rm g},A)=S_{k}({\rm g})+\frac{1}{4\pi}\int \hs{-1mm}d^2\xi
Tr(A_+\pr_-{\rm g}{\rm g}^{-1}\hs{-1mm}-A_-g^{-1}\pr_+{\rm g}+
A_+{\rm g}A_-{\rm g}^{-1}-\hs{-1mm}A_-A_+),\label{actioni}
\ee
where the gauge fields may be parametrized as
\be
A_+=\pr_+{\rm h}{\rm h}^{-1} \hs{10mm} A_-=\pr_-\ti{{\rm h}}
\ti{{\rm h}}^{-1},
\ee
and ${\rm h},\ \ti{{\rm h}}$ are elements of $H$.
The partition function may
now, after integrating out the gauge fields, be written as
\cite{Goddard-Olive86},\cite{Karabali-Park-Schnitzer-Yang89}
\be
Z=&\int&[d{\rm g}][d\tilde{{\rm h}}][db_+][db_-][dc_+][dc_-]
exp[-kS_{k}({\rm g})]
exp[-(-k-2c_H)S_{- k-2{c}_H}(\tilde{{\rm h}})]\nonumber\\
&\times& exp[-Tr\int d^2\xi (b_+\pr_-c_++b_-\pr_+c_-)].\label{part}
\ee
We thus find that the gauged WZNW models factorize into
three parts, the original model of group $G$ and level $k$,
another WZNW model of group $H$ and level $-k-2c_H$ and a
ghost sector. \\
\ind The conformal anomaly of the Sugawara-type stress-energy
tensor of the gauged WZNW model
coincides with the GKO coset conformal anomaly,
as noted in \cite{Karabali-Park-Schnitzer-Yang89}.  The total
action corresponding to the partition function eq.(\ref{part})
is invariant under the BRST transformation see e.g.
\cite{Karabali-Schnitzer84}, \cite{Bastianelli91}.
The BRST charge of the holomorphic sector is found to
be
\be
Q=\oint\frac{dz}{2i\pi}\left[:c_a(z)(J^a(z)+\jt^a(z)):-\frac{i}{2}
f^{ad}_{\ \
 e}
:c_a(z)c_d(z)b^e(z):\right].\label{brstq}
\ee
Using the BRST invariance, we find that the stress-energy tensor of
the GKO coset construction $T^{GKO}=T^G-T^H$
coincides with the
stress-energy tensor of the gauged WZNW model
$T^{tot}=T^G+T^{\ti H}+T^{gh}$ up to
commutators with the BRST charge, i.e.
\be
T^{tot}(z)=T^{GKO}(z)+\frac{1}{k+c_H}\left[Q,:b_a(z)
\left( J^a(z)-\jt^a(z)
\right):\right].\label{emq}
\ee
We then find from the BRST physicality condition
\be
Q|{\rm phys}\rangle=0\label{brsteq}
\ee
that the stress-energy tensors of the two models will coincide on
the physical sub-space. \\
\ind In \cite{Hwang-Rhedin93} the BRST condition eq.(\ref{brsteq})
was analyzed, or rather the BRST cohomology was computed
with the restriction that we have integrable
representations for the currents of the $G$ sector.
This integrability condition
may be expressed as
\be
\al\cdot\mu\geq 0 \hs{6mm} \forall \al \in\De^s_g, \label{intrgi}
\ee
where $\De^s_g$ is the set of simple roots of $g$, and
\be
\hat{\al}\cdot\mu\leq\frac{k}{2}, \label{intrgii}
\ee
where $\hat{\al}$ is the highest root and $\mu$ is
the highest weight of the vacuum
representation, \cite{Goddard-Olive86}.  It was found in
\cite{Hwang-Rhedin93}
that the GKO coset condition eq.(\ref{coset}) coincides with the
BRST condition eq.(\ref{brsteq}) when we have integrable
representations for the $G$ sector and no null vectors in the
$\ti{H}$ sector, (i.e., no states that have vanishing inner
products with all other states). \\
\ind To make the rest of the paper more accessible we give the
affine Lie algebra in the Cartan-Weyl basis,
\be
[J_m^i,J_n^j]&=&\frac{k}{2} m\delta^{ij}\delta_{m,-n}
\nonumber \\
{}~[J_m^i, J_n^\alpha ]&=& \alpha ^iJ_{m+n}^\alpha
\nonumber\\
 ~[J_m^\al,J_n^\beta] &=& \left\{\begin{array}{ll} \epsilon(\al,\beta)
J_{m+n}
^{\al+\beta}\hs{30mm} &
\mbox{if $\alpha +\beta$ is a root}
 \\
\frac{1}{\al^2}(\al_iJ^i_{m+n}+\frac{k}{2}m\del_{m,-n}) \hs{6mm} &
\mbox{ if
 $\al=-\beta$ } \\
 0 \hs{48mm} & \mbox{otherwise.}\label{cwbasis}
\end{array}
\right.
\ee
We also denote $\De_g^+=$ the set of positive roots of $g$.
We introduce vacuum states of the
three sectors defined to obey
\be
& & J_n^A|0;R\ran=0 \hs{10mm} n>0 \ \ {\rm or} \ \ n=0 \
{\rm and} \ A\in\De^+_g \nn \\
& & \ti{J}_n^a|\ti{0};\ti{R}\ran=0 \hs{11mm} n>0
\ \ {\rm or} \ \ n=0 \ {\rm and} \ a\in\De^+_h \nn \\
& & b_n^a|0\ran_{gh}=0 \hs{13mm} n\geq0 \nn \\
& & c_n^a|0\ran_{gh}=0 \hs{13mm} n\geq1 .
\ee
An arbitrary state is then given by
$|s\ran=|s_G\ran\times|s_{\ti{H}}\ran\times|s_{gh}\ran$
where
\be
| s_G\rangle =\sum_R\prod_{A,n}J^A_{-n}|0;R\rangle,\hs{5mm}|
s_{\widetilde{H}}
\rangle =\sum_{\tilde R}\prod_{a,n}\jt_{-n}^a| \tilde{0};
\widetilde{R}\rangle,
 \hs{25mm} \label{state}
\ee
and $|s_{gh}\ran$ is a sum over states of the form
\be
\prod_{a_1,n_1}\prod_{a_2,n_2}b_{-n_1}^{a_1}c_{-n_2}^{a_2}| 0\rangle_{gh}.
\ee
The state $|0\rangle_{gh}$ is the $SL(2,{\bf R})$ invariant ghost vacuum,
which is required to be annihilated by $L^{gh}_n$ for $n=0,\pm 1$.
The state $|0;R\rangle$ is a highest weight primary with respect
to the affine Lie algebra $\hat{g}$, and transforms in some
representation $R$ of the Lie algebra $g$. Likewise for the
auxiliary sector. This means that the product in (\ref{state}) is taken
over $n>0$ or $n=0$ and negative roots.
For integrable representations the $G$ sector may always be
decomposed into states of the type \cite{Kac-Peterson84}
\be
| s_G\rangle=\sum_{\Phi}\prod_{a,n}J^a_{-n}|\Phi\rangle\label{decomp},
\ee
where $|\Phi\ran$ are primary with respect to the currents
of the subgroup $H$. \\
\ind It is convenient to first study the problem on the
relative state space \cite{Frenkel-Garland-Zuckerman86} defined as
\be
b_{0,i}|s\rangle=0\hs{20mm}i=1,\ldots ,r_h,\label{relspace}
\ee
where $r_h$ is the rank of the Lie algebra $h$.
This means that we can split the BRST charge as
\be
Q=\hat{Q}+M_ib^i_0+c_{0,i}J^{tot,i}_0\label{brstdecomp}.
\ee
Since $J_m^{i,tot}$ is a BRST commutator we find that
states in the relative cohomology must have zero eigenvalue of
$J_0^{i,tot}$
and that on the relative state space the $Q$ cohomology
coincides with the $\hat{Q}$ cohomology. \\

When we analyze the relative cohomology we will find
\cite{Hwang-Rhedin93} that there are no states at ghost number
different from zero.
To be more specific, if we restrict the choices of representations
of the $\ti{H}$ sector in an analogous manner to eq.(\ref{intrgi}) and
eq.(\ref{intrgii}) namely
\be
\al\cdot(\ti{\mu}+\rho_h)\leq0, \label{tirestri}
\ee
and
\be
\hat{\al}\cdot\ti{\mu}\geq\frac{\ti{k}}{2}+1, \label{tirestrii}
\ee
where $\al\in\De^s_h$, $\hat{\al}$ is the highest root and
$\ti{k}=-k-2c_H$, then the relative cohomology contains
only states of ghost number zero.
Furthermore, those states have no
$\ti{J}^a$ excitations and they satisfy
\be
J^a_n|\phi\ran=b^a_n|\phi\ran=c^a_n|\phi\ran=0
\ee
for $n>0$ or $n=0$ and $a\in\De_h^+$. We should also
remark that this choice of representations (\ref{tirestri}),
(\ref{tirestrii}) is the one that excludes all null vectors in the
auxiliary sector. It also means that $-\ti{\mu}-\rho_h$ is an
integrable weight, a fact which will be important in the next section.
This is another way of finding the appropriate representation restrictions
(\ref{tirestri}), (\ref{tirestrii}). For the case of a $U(1)$ group, we
have that the eigenvalue of the Cartan generator of the auxiliary
sector take values in ${\cal Z}/2$. \\
\ind We thus conclude that for this choice of representation the
relative cohomology is completely specified. \\
\ind Turning to the cohomology of the full space we will only comment that
the absolute cohomology contains more states as can be seen from the
$2^{r_h}$ degeneracy of the ghost vacua. This degeneracy is removed
by requiring the physical states to be in the relative state-space.

\section{Characters}

We are now ready to define the BRST invariant character
of the gauged WZNW model.
In general the affine character is defined as
\be
\chi(\tau,\theta)=Tr\left(e^{2i\pi\tau (L_0^G+L_0^{\ti{H}}+L_0^{gh})}
e^{i(\theta_iJ^{tot,i}_0+\theta_{I^{\prime}}J_0^{I^{\prime}}})
(-1)^{N_{gh}}\right),\label{char}
\ee
where we have omitted the usual factor $e^{-2i\pi\tau c/24}$ for
convenience.  Here $i$ take labels in $h$ and $I'$ in $g/h$. We assume
the existence of a basis such that this decomposition is possible.
Since the physical states live on the
relative state space, we do not include a summation over
ghost vacua in the trace. We exclude this part of the trace to
remove the
ghost vacuum degeneracy.  On the physical state space we have in
addition that $J^{tot,i}_0|{\rm phys}\ran=0$. This follows from
the definition of
the relative state space $b^i_0|\phi\ran=0$, the BRST condition
$Q|{\rm phys}\ran=0$ and the fact that $J^{tot,i}_0=\{Q,b^i_0\}$.
To only have physical degrees of freedom propagating we must,
therefore, implement
this condition in the character.  This can be done
by integrating over the $\th$-components
associated with $J^{tot,i}_0$, i.e. we define
\be
\chi^{G/H}(\tau,\th)=\int\prod_i\frac{d\theta_i}{2\pi}
Tr\left(e^{2i\pi\tau(L_0^G+
L_0^{\ti{H}}+L_0^{gh})}e^{i(\theta_iJ_0^{tot,i}+\theta_{I^{\prime}}
J_0^{I^{\prime}})}(-1)^{N_{gh}}\right).\label{chari}
\ee
This definition of the character now respects the BRST symmetry of
the model,
and we are then assured that only states in the coset will propagate.
In what follows we will instead of $\tau$
use $q\equiv e^{2\pi i\tau}$. In order to make the formulae
more transparent we choose to
display characters only for simply-laced algebras. \\
\ind The BRST invariant  character (\ref{chari}), is a product of
three different terms $\chi^G$,
$\chi^{\tilde{H}}$ and $\chi^{gh}$. $\chi^G$ is given by the
Kac-Weyl formula \cite{Kac74},
\be
\chi^G(q,\th)&=&\sum_{t\in \La ^{\nu}_r}q^{\frac{(\la+\rho/2+t)^2-
\rho^2/4}{k+c_G}}\sum_{\si\in W(g)}\ep(\si)
e^{i(\si(\la+\rho/2+t)-\rho/2)\cdot\th}R_G^{-1}(q,\th) \label{weylkac},
\ee
where $\La ^{\nu}_r$ denotes a lattice which is
spanned by $t=n(k+c_G)\al\psi^2/\al^2$ where $\al\in\De^s_g$,
$n\in{\cal Z}$ and $\psi$ is a long root, ( which we have chosen to
normalize as $\psi^2=1$ ). $t$ is thus
proportional to the co-roots of $g$.
$\si$ are elements of the Weyl group $W(g)$, and
\be
R_G(q,\th)=\prod_{n=1}^{\infty}(1-q^n)^{r_g}\prod_{\al\in\De_g^+}(1-
q^ne^{i\al\cdot\th})(1-q^{n-1}e^{-i\al\cdot\th}). \label{rg}
\ee
$r_g=$rank$(g)$ and $\De_g^+$ is the set of positive roots
of $g$. \\
\ind $\chi^{\tilde{H}}$ is straightforward to determine, since we
know that there are no null vectors in the auxiliary sector.
A moments reflection yields us
\be
\chi^{\tilde{H}^+}(q,\theta) =
e^{i\theta\cdot\tilde{\mu}}q^{-\frac{\tilde{\mu}\cdot
(\tilde{\mu}+\rho)}{k+c_H}
}R^{-1}_H(q,\theta)\label{chart},
\ee
for highest weight representations.
The ghost
contribution may be constructed for each pair $b_n^a$ and
$c_n^a$.  They contribute as does conformal ghosts, see eg.
\cite{Ginsparg88}, with the exception of the twist according to
the eigenvalue of $J_0^{gh,i}$.  For the zero modes $b_0^{\al}$,
$c_0^{\al}$ we
find $e^{i\rho\cdot\theta}\prod_{\alpha>0}(1-e^{-i\alpha\theta})^{2}$.
This leaves us with
\be
\chi^{gh}(q,\theta)=e^{i\rho\cdot\theta}R_H^2(q,\theta)\label{ghchar}
\ee
for the ghost sector. \\

In ref. \cite{Hwang-Rhedin94} it is shown that the BRST invariant character
coincides with the usual definition of a branching function,
\be
\chi^G_{\La}=\sum_{\la}\chi^{G/H}_{\La,\la}\chi^H_{\la}, \label{branch}
\ee
where we take the sum over integrable highest weights $\la$ of $H$. \\
\ind Furthermore, in reference \cite{Hwang-Rhedin94}, the branching function of
$G/(U(1))^x$ models are calculated, for $1\leq x\leq r_g$ where
$r_g$ is the rank of the Lie algebra $g$.
A general formula for string functions for any group $G$ is also
displayed.  \\
\ind In \cite{Hwang-Rhedin93} the coincidence between the BRST
invariant character and conventional character
calculations were shown for the parafermion model $SU(2)_k/U(1)$
previously obtained in \cite{Distler-Qiu90},
and minimal Virasoro models $SU(2)_k\times SU(2)_1/SU(2)_{k+1}$
known from \cite{Goddard-Kent-Olive86}.

Here we will first compute the characters
of $G/H$ cosets such that
rank$(G/H)=0$ as well as $(G\times G)/G$ models.
Special cases have occurred in several publications.
Apart from those mentioned above we have for instance
$SU_k(N)\times SU_l(N)/SU_{k+l}(N)$ in \cite{Ravanini88},
\cite{Kastor-Martinec-Qiu88}
and \cite{Bagger-Nemeschansky-Yankielowicz88}. \\
\ind Furthermore, all these
models are all computed in ref. \cite{Bouwknegt-McCarthy-Pilch91}
but only under
certain assumptions.  The $(G\times G)/G$ model is also calculated
in ref. \cite{Christe-Ravanini89} using a different method
than \cite{Bouwknegt-McCarthy-Pilch91}.  Characters
of these kinds are also computed in \cite{Huito-Nemeschansky-Yankielowicz90} in
a
general context.  Reference \cite{Huito-Nemeschansky-Yankielowicz90} is,
however,
restricted to a few explicit examples.  \\
\ind Both \cite{Bouwknegt-McCarthy-Pilch91} and \cite{Christe-Ravanini89}
uses free field techniques,
although they choose different approaches.
Bouwknegt, McCarthy and Pilch, \cite{Bouwknegt-McCarthy-Pilch91},
uses free field Fock spaces
to obtain resolutions of heighest weight modules for $G/H$
coset models.  This is done using techniques of double
complexes.  Firstly one utilizes an operator first introduced by Felder
in the context of minimal models \cite{Felder89}, and
Bernard and Felder \cite{Bernard-Felder90}, for $SL(2,{\bf R})$,
and secondly a BRST operator.
The assumption made is
that the cohomologies of the Felder-like as well as the
BRST operator are constrained to ghost number zero.
This has so far only been proven for the case of
minimal models \cite{Felder89}, and for $G=Sl(2,{\bf R})$
\cite{Bernard-Felder90}, for
the Felder-like operator. The result of the cohomology of the
BRST operator is later proven in \cite{Bouwknegt-McCarthy-Pilch92:2}
but only under the
the assumption that the Felder-like operator cohomology
is non-trivial only for ghost number zero. \\
\ind Christie and Ravaninni \cite{Christe-Ravanini89} decompose
$G_{k_1}\times G_{k_2}/G_{k_1+k_2}$ in parafermions and free fields.
They are then able to write down a branching function
in terms of string functions \cite{Kac-Peterson84}.
Those branching functions are proven to be in an explicit form in
\cite{Hwang-Rhedin93}.  \\

\noindent {\bf{Characters of {\boldmath $G/H$} for rank{\boldmath
$(G/H)=$}zero.}} \\
We first assume that $H$ is a direct sum of simple
subgroups $H_a$. For the $G$ sector we find the character,
(\ref{weylkac}),(\ref{rg})
\be
\chi^G_{\la}(q,\th)&=&\sum_{t\in \La ^{\nu}_r}q^{\frac{(\la+\rho/2+t)^2-
\rho^2/4}{k+c_G}}\sum_{\si\in W(g)}\ep(\si)
e^{i(\si(\la+\rho/2+t)-\rho/2)\cdot\th}\nn \\
&\cdot&\lef(\prod_{n=1}^{\infty}(1-q^n)^{r_g}\prod_{\al\in\De_g^+}(1-
q^ne^{i\al\cdot\th})(1-q^{n-1}e^{-i\al\cdot\th})\rig)^{-1}.
\ee
For each subgroup we find from eq.(\ref{chart}) and eq.(\ref{ghchar})
the $\ti{H}$ character and the ghost contribution to be
\be
\chi^{\ti{H}}_{\ti{\mu}_a}(q,\th)=e^{i\th_a\cdot\ti{\mu}_a}
q^{\frac{-\ti{\mu}_a\cdot
(\ti{\mu}_a+\rho_a)}
{k+c_{H_a}}}\lef(\prod_{n=1}^{\infty}
\prod_{\al\in\De_{h_a}^+}(1-q^n)^{r_{h_a}}(1-
q^ne^{i\al\cdot\th_a})(1-q^{n-1}e^{-i\al\cdot\th_a})\rig)^{-1}
\ee
and
\be
\chi^{gh_a}(q,\th)=e^{i\th_a\cdot\rho_a}\lef(\prod_{n=1}^{\infty}
\prod_{\al\in\De_{h_a}^+}(1-q^n)^{r_{h_a}}(1-
q^ne^{i\al\cdot\th_a})(1-q^{n-1}e^{-i\al\cdot\th_a})\rig)^2
\ee
respectively. The range of permissible weights $\ti{\mu}_a$ is given
from the choice of representations (\ref{tirestri}), (\ref{tirestrii})
and we have that $-\ti{\mu}_a-\rho_a$ take values in
the set of integrable representations. \\
\ind We now insert those ingredients into the integral
\be
\chi^{G/H}(\tau,\th)=\int\prod_i\frac{d\theta_i}{2\pi}\chi^G\chi^{\ti{H}}
\chi^{gh}.
\ee
Under the assumption of the embedding of $h$ in $g$,
i.e. that there exists a basis for $g$ such that the Cartan
sub-algebra of $h$ is just a subset of the Cartan sub-algebra
of $g$,
we have that $\al_a\cdot\th_a=\al\cdot\th$ for some
$\al_a\in\De^+_{h_a}$ and $\al\in\De^+_g$.
In order to perform the integration, we use the identity due to
Thorn \cite{Thorn89}
\be
&\ &\lef(\prod_{n=1}^{\infty}\prod_{\al\in\De_g^+}(1-
q^ne^{i\al\cdot\th})(1-q^{n-1}e^{-i\al\cdot\th})\rig)^{-1}
\nn \\
&=&\prod_{n=1}^{\infty}(1-q^n)^{-2|\De^+_g|}
\prod_{j=1}^{|\De^+_g|}\sum_{p_j\in{\cal Z}}\sum_{s_j =0}
^{\infty}(-1)^{s_j}q^{\frac{1}{2}(s_j-p_j+1/2)^2
-\frac{1}{2}(p_j-1/2)^2}e^{ip_j\th\cdot\al_j}\label{thornid}
\ee
where $|\De^+_g|$ denotes the number of positive roots of $g$.
Performing the
integration we will find
\be
\chi^{G/H}(q,\th)&=&\frac{1}{\prod_{n=1}^{\infty}(1-q^n)^{2|\De^+_
{g/h}|}
}\sum_{\si\in W(g)}\sum_{t\in \La ^{\nu}_{r,g}}\nn \\ &\cdot&
\ep(\si)q^{\frac{(\la+\rho/2
+t)^2-\rho^2/4}{k+c_G}}q^{-\sum_a\frac{\ti{\mu}_a\cdot
(\ti{\mu}_a+\rho_a)}{k+c_{H_a}}} \nn \\
&\cdot&\prod_{j=1}^{|\De^+_
{g/h}|}\sum_{p_j\in{\cal Z}}^{\ \ \ \ \ \prime}\sum_{s_j=0}^{\infty}
(-1)^{s_j}q^{\frac{1}{2}(s_j-p_j+1/2)^2
-\frac{1}{2}(p_j-1/2)^2}.
\ee
The restriction on
the sum of $p_j$, indicated by the prime on the sum, should be taken as
\be
\si(\la+\frac{\rho}{2}+t)-\frac{\rho}{2}+\ti{\mu}+\rho+
\sum_{j=1}^{|\De^+_{g/h}|}p_j\al_j=0.
\ee
Here $\ti{\mu}$ is an $r_g$ dimensional vector such that
$\sum_a\ti{\mu}_a\cdot\th_a=\ti{\mu}\cdot\th$. \\
\ind This is in agreement with the result of reference
\cite{Bouwknegt-McCarthy-Pilch91}.
Since our result is derived without any assumptions whatsoever
this is a verification of the result of
Bouwknegt, McCarthy and Pilch \cite{Bouwknegt-McCarthy-Pilch91}.
For the choice $G=SU(3)$ and $H=SU(2)\times U(1)$ this is also
found to agree with the result of Huito, Nemeshansky and
Yankielowicz \cite{Huito-Nemeschansky-Yankielowicz90}. \\

\noindent {\bf{{\boldmath $G_{k_1}\times G_{k_2}/G_{k_1+k_2}$} models. }}
\\
In those models one usually chooses to divide out the
diagonal subgroup, a prescription which we also chose to follow. \\
\ind Our starting point is as usual the BRST invariant character
\be
\chi^{G/H}(\tau,\th)=\int\prod_i\frac{d\theta_i}{2\pi}\chi^G\chi^{\ti{H}}
\chi^{gh}.
\ee
The characters of the two groups $G_{k_1}$ and  $G_{k_2}$ are given
from
\be
\chi^G_{\la}(q,\th)&=&\sum_{t\in \La ^{\nu}_r}q^{\frac{(\la+\rho/2+t)^2-
\rho^2/4}{k+c_G}}\sum_{\si\in W(g)}\ep(\si)
e^{i(\si(\la+\rho/2+t)-\rho/2)\cdot\th}\nn \\
&\cdot&\lef(\prod_{n=1}^{\infty}(1-q^n)^{r_g}\prod_{\al\in\De_g^+}(1-
q^ne^{i\al\cdot\th})(1-q^{n-1}e^{-i\al\cdot\th})\rig)^{-1},
\ee
with labels 1 and 2 respectively i.e.
$\chi^G_{\la_1,\la_2}=\chi^G_{\la_1}\chi^G_{\la_2}$.  \\
\ind For the auxiliary
sector, in which the range of weights $\ti{\la}_3$ are
restricted according to the our choice (\ref{tirestri}), (\ref{tirestrii}),
we find the character
\be
\chi^{G}_{\ti{\la}_3}(q,\th)=e^{i\th\cdot\ti{\la}_3}q^{\frac{-\ti{\la}
_3\cdot(\ti{\la}_3+\rho)}
{k_1+k_2+c_G}}\lef(\prod_{n=1}^{\infty}
\prod_{\al\in\De_g^+}(1-q^n)^{r_g}(1-
q^ne^{i\al\cdot\th})(1-q^{n-1}e^{-i\al\cdot\th})\rig)^{-1}.
\ee
In order to compare our result with other authors, we note that
$-\ti{\la}_3-\rho$ is an integrable weight. \\
\ind As for the ghosts we have
\be
\chi^{gh}(q,\th)=e^{i\th\cdot\rho}\lef(\prod_{n=1}^{\infty}
\prod_{\al\in\De_g^+}(1-q^n)^{r_g}(1-
q^ne^{i\al\cdot\th})(1-q^{n-1}e^{-i\al\cdot\th})\rig)^2.
\ee
The integration is carried out using identity (\ref{thornid}), and
the result becomes
\be
\chi^{G/H}_{\la_1,\la_2,\ti{\la}_3}(q,\th)&=&
\frac{1}{\prod_{n=1}^{\infty}(1-q^n)^{2|\De^+_g|+r_g}
}\sum_{\si_1\in W(g)}\sum_{\si_2\in W(g)}\sum_{t_1 \in \La ^{\nu}_{r,g}}
\sum_{t_2 \in \La ^{\nu}_{r,g}}\nn \\ &\cdot&
\ep(\si_1)\ep(\si_2)q^{\frac{(\la_1+\rho/2
+t_1)^2-\rho^2/4}{k_1+c_G}}q^{\frac{(\la_2+\rho/2
+t_2)^2-\rho^2/4}{k_2+c_G}}
q^{-\frac{\ti{\la}_3\cdot(\ti{\la}_3+\rho)}{k_1+k_2+c_G}} \nn \\
&\cdot&\prod_{j=1}^{|\De^+_
g|}\sum_{p_j\in{\cal Z}}^{\ \ \ \ \ \prime}\sum_{s_j=0}^{\infty}
(-1)^{s_j}q^{\frac{1}{2}(s_j-p_j+1/2)^2 -\frac{1}{2}(p_j-1/2)^2}.
\label{gsqr}
\ee
The restriction of the primed sum should be taken as
\be
\si_1(\la_1+\frac{\rho}{2}+t_1)-\frac{\rho}{2}+
\si_2(\la_2+\frac{\rho}{2}+t_2)-
\frac{\rho}{2}+\ti{\la}_3+\rho+\sum_{j=1}^{|\De^+_g|}p_j\al_j=0.
\label{primrestr}
\ee
\ind In eq.(\ref{gsqr}) we have suppressed the non-diagonal $\th$
behavior partly for clarity, and partly since this is standard
in literature. In general, $\chi^{G/H}$ of course depends on
the $\th$ components that we do not integrate over, but we may
choose to write the character in the point of moduli-space
where the $\th$ dependence vanishes.  In analyzing the modular
properties things would, however, be more transparent if the
$\th$ dependence was kept intact. \\
\ind Our result (\ref{gsqr}), (\ref{primrestr}) essentially agree with
the result of ref. \cite{Bouwknegt-McCarthy-Pilch91}.
There is a difference of a constant factor
\be
\frac{\rho^2}{4}\frac{k_1}{(k_2+c_G)(k_1+k_2+c_G)}.
\ee
\ind For the case of $G=SU(2)$ this is found to agree with previous
calculations \cite{Kastor-Martinec-Qiu88} and
\cite{Bagger-Nemeschansky-Yankielowicz88}. These models are also
discussed in \cite{Ravanini88} but no explicit formula
for the branching function is given. \\
\ind Our result should also agree with \cite{Christe-Ravanini89}
where the branching
functions are given in terms of string functions. (An explicit
form of those string functions is proven in \cite{Hwang-Rhedin93}.)
\\

\noindent{\bf{Not previously considered models.}}
\\
We will now compute the obvious
generalization of the $G_{k_1}\times G_{k_2}/G_{k_1+k_2}$ case
namely $\prod^{m}_{a=1}G_{k_a}/G_{\sum^{m}_{a=1}k_a}$ models.
This construction is the natural building block of any model
of the type $G^m/G^n$, ( the notation should be obvious ),
since $G/G$ is essentially unity. \\
\ind We will also consider $G_{k_1}\times H_{k_2}/H_{k_1+k_2}$,
where $H$ is a subgroup of $G$.
For the case of $SU_k(2)\times U(1)/U(1)$ this has been
calculated in \cite{Ravanini-Yang87}. \\

\noindent{\bf{ The {\boldmath $\prod^{m}_{a=1}G_{k_a}/G_
{\sum^{m}_{a=1}k_a}$} models.}} \\
We will here follow the prescription given for the case
$m=2$ given above and choose to divide out the diagonal
group. For the numerator we here use $m$ copies of
\be
\chi^G_{\la_a}(q,\th)&=&\sum_{t_a\in \La ^{\nu}_r}
q^{\frac{(\la_a+\rho/2+t_a)^2-
\rho^2/4}{k_a+c_G}}\sum_{\si_a\in W(g)}\ep(\si_a)
e^{i(\si_a(\la_a+\rho/2+t_a)-\rho/2)\cdot\th}\nn \\
&\cdot&\lef(\prod_{n=1}^{\infty}(1-q^n)^{r_g}\prod_{\al\in\De_g^+}(1-
q^ne^{i\al\cdot\th})(1-q^{n-1}e^{-i\al\cdot\th})\rig)^{-1},
\ee
where of course $1\leq a\leq m$. \\
\ind The character for the auxiliary sector is found to be
\be
\chi^G_{\ti{\mu}}(q,\th)=e^{i\th\cdot\ti{\mu}}
q^{\frac{-\ti{\mu}
\cdot(\ti{\mu}+\rho)}
{K+c_G}}\lef(\prod_{n=1}^{\infty}
\prod_{\al\in\De_g^+}(1-q^n)^{r_g}(1-
q^ne^{i\al\cdot\th})(1-q^{n-1}e^{-i\al\cdot\th})\rig)^{-1},
\ee
where we have introduced
\be
K\equiv\sum_{a=1}^m k_a. \label{manyk}
\ee
The set of $\ti{H}$ weights $\ti{\mu}$ is restricted according
to our choice (\ref{tirestri}, (\ref{tirestrii}) which means
that $-\ti{\mu}-\rho$ take integrable representations.
We also find for the ghosts
\be
\chi^{gh}(q,\th)=e^{i\th\cdot\rho}\lef(\prod_{n=1}^{\infty}
\prod_{\al\in\De_g^+}(1-q^n)^{r_g}(1-
q^ne^{i\al\cdot\th})(1-q^{n-1}e^{-i\al\cdot\th})\rig)^2.
\ee
\ind When we perform the integration
\be
\chi^{G/H}(\tau,\th)=\int\prod_i\frac{d\theta_i}{2\pi}\chi^G\chi^{\ti{H}}
\chi^{gh}.
\ee
we may as usual benefit from the identity due to Thorn \cite{Thorn89}
\be
&\ &\lef(\prod_{n=1}^{\infty}\prod_{\al\in\De_g^+}(1-
q^ne^{i\al\cdot\th})(1-q^{n-1}e^{-i\al\cdot\th})\rig)^{-1}
\nn \\
&=&\prod_{n=1}^{\infty}(1-q^n)^{-2|\De^+_g|}
\prod_{j=1}^{|\De^+_g|}\sum_{p_j\in{\cal Z}}\sum_{s_j =0}
^{\infty}(-1)^{s_j}q^{\frac{1}{2}(s_j-p_j+1/2)^2
-\frac{1}{2}(p_j-1/2)^2}e^{ip_j\th\cdot\al_j}
\ee
and accordingly find
\be
\chi^{G/H}_{\la_a,\ti{\mu}}(q,\th)&=&\frac{1}{\prod_{n=1}^{\infty}
(1-q^n)^{2
(m-1)|\De^+_g|+(m-1)r_g}
}\sum_{\si_1\in W(g)}...\sum_{\si_m\in W(g)}\sum_{t_1\in \La ^
{\nu}_{r,g}}...
\sum_{t_m\in \La ^{\nu}_{r,g}}\nn \\
&\cdot&\ep(\si_1)...\ep(\si_m)
q^{\sum_{a=1}^m\frac{(\si_a(\la_a+\rho/2
+t_a))^2-\rho^2/4}{k_a+c_G}}q^{-\frac{\ti{\mu}\cdot(\ti{\mu}+\rho)}
{K+c_G}} \nn \\
&\cdot&
\prod_{a=1}^{m-1}\prod_{j_a=1}^{|\De^+_
g|}\sum_{p_{j_a}\in{\cal Z}}^{\ \ \ \ \ \prime}\sum_{s_{j_a}=0}^{\infty}
(-1)^{s_{j_a}}q^{\frac{1}{2}(s_{j_a}-p_{j_a}+1/2)^2
-\frac{1}{2}(p_{j_a}-1/2)^2}, \label{manyg}
\ee
where again $K$ is given by (\ref{manyk}).
The restriction of the sum of $p_j$'s is
\be
\sum_{a=1}^m(\si_a(\la_a+\frac{\rho}{2}+t_a))-m\frac{\rho}{2}+
\ti{\mu}+\rho+\sum_{a=1}^{m-1}\sum_{j_a=1}^{|\De^+_g|}p_{j_a}\al_{j_a}=0.
\ee
As in the case of $G^2/G$ we have chosen to suppress the
non-diagonal $\th$ behavior in eq.(\ref{manyg}). \\
\noindent{\bf{ {\boldmath $G_{k_1}\times H_{k_2}/H_{k_1+k_2}$} where
{\boldmath $H\subset G$}.}} \\
We take the embedding of $H_{k_1+k_2}$ to be the diagonal one. To make
the formulae simpler we take $H$ to be simple. The generalization
to non-simple $H$ is straightforward.
The character of $G_{k_1}$
is given by the Kac-Weyl formula (\ref{weylkac})
\be
\chi^G_{\La}(q,\th)=\sum_{t_g\in \La ^{\nu}_{r,g}}q^
{\frac{(\La+\rho/2+t_g)^2-
\rho^2/4}{k_1+c_G}}\sum_{\si_g\in W(g)}\ep(\si)
e^{i(\si(\La+\rho/2+t_g)-\rho/2)\cdot\th}R_G^{-1}(q,\th),
\label{weylkacagain}
\ee
where we as usual have
\be
R_G(q,\th)=\prod_{n=1}^{\infty}(1-q^n)^{r_g}\prod_{\al\in\De_g^+}(1-
q^ne^{i\al\cdot\th})(1-q^{n-1}e^{-i\al\cdot\th}). \label{rgagain}
\ee
For $H_{k_2}$ we also find
from (\ref{weylkacagain})
\be
\chi^H_{\la}(q,\bar{\theta})&=&\sum_{t_h\in \La ^{\nu}_{r,h}}
q^{\frac{(\la+\rho_h/2+t_h)^2-
\rho_h^2/4}{k_2+c_H}}\sum_{\si_h\in W(h)}\ep(\si_h)
e^{i(\si_h(\la+\rho_h/2+t)-\rho_h/2)\cdot\bar{\th}}\nn \\
&\cdot&\lef(\prod_{n=1}^{\infty}(1-q^n)^{r_h}\prod_{\al\in\De_h^+}(1-
q^ne^{i\al\cdot\bar{\th}})(1-q^{n-1}e^{-i\al\cdot\bar{\th}})\rig)^{-1}.
\ee
For the auxiliary sector the set of allowed representation is
restricted such that we have integrable representations for
$-\ti{\mu}-\rho_h$, cf.
equations (\ref{tirestri}) and (\ref{tirestrii}).
This gives us the character
\be
\chi^{\tilde{H}}_{\ti{\mu}}(q,\ti{\theta}) =
e^{i\ti{\theta}\cdot\tilde{\mu}}q^{-\frac
{\tilde{\mu}\cdot(\tilde{\mu}+\rho_h)}{k_1+k_2+c_H
}}R^{-1}_H(q,\ti{\theta}),
\ee
for the auxiliary sector, and we find for the ghost sector
\be
\chi^{gh}(q,\ti{\theta})=e^{i\rho_h\cdot\ti{\theta}}R_H^2(\tau,\ti{\theta}).
\ee
$R_H$ is of course given by (\ref{rgagain}) with labels $H$ and
$h$ instead of $G$ and $g$.
With $h$ embedded in $g$ under the usual assumption we can take
$\al_h\cdot\ti{\th}=\al\cdot\th$, for some $\al_h\in\De^+_h$ and
$\al\in\De^+_g$. We will in the following suppress the non-diagonal
$\th$ behavior. We now split $\th$ into a direct sum of the $r_h$ first
components $\th_h$, and the last $r_g-r_h$ components $\th'$, that is
$\th=\th_h\oplus\th'$. We will then integrate over $\th_h$ while the
remaining diagonal $\th$ behavior will be displayed by $\th'$.
We should then perform the integral
\be
\chi^{G/H}(\tau,\th)=\int\frac{d\theta_h}{2\pi}\chi^G\chi^{\ti{H}}
\chi^{gh}.
\ee
Using Thorn's trick, (\ref{thornid}) we find the character to be
\be
\chi^{G/H}_{\La,\la,\ti{\mu}}(q,\th')&=&
\frac{1}{\prod_{n=1}^{\infty}(1-q)^{r_g+2|\De^+_g|}}
\sum_{\si_g\in W(g)}\sum_{\si_h\in W(h)}
\sum_{t_g\in \La ^{\nu}_{r,g}}\sum_{t_h\in \La ^{\nu}_{r,h}}
\ep(\si_g)\ep(\si_h)\nn \\
&\cdot&q^{\frac{(\La+\rho_g/2+t_g)^2-
\rho_g^2/4}{k_1+c_G}}q^{\frac{(\la+\rho_h/2+t_h)^2-\rho_h^2/4}{k_2+c_H}}
q^{-\frac{\ti{\mu}\cdot(\ti{\mu}+\rho_h)}{k_1+k_2+c_H}}
e^{i(\si_g(\La+\rho_g/2+t_g)-\rho_g/2)\cdot\th'} \nn \\
&\cdot&\prod_{j=1}^{|\De^+_g|}\sum_{p_j\in{\cal Z}}^{\ \ \ \ \ \prime}
\sum_{s_j=0}^{\infty}
(-1)^{s_j}q^{\frac{1}{2}(s_j-p_j+1/2)^2
-\frac{1}{2}(p_j-1/2)^2}e^{ip_j\al_j\cdot\th'} ,
\ee
where the restriction on the sum over $p_j$'s is
\be
\lef(\si_g(\La+\frac{\rho_g}{2}+t_g)-\frac{\rho_g}{2}+
\si_h(\la+\frac{\rho_h}{2}+t_h)-
\frac{\rho_h}{2}+
\ti{\mu}+\rho_h+\sum_{j=1}^{|\De^+_g|}p_j\al_j\rig)_i=0. \hs{5mm}
\ee
The index $i$ on the bracket is in the range $1\leq i\leq r_h$
since we only require the first $r_h$ components of this vector
to vanish.

\section{Concluding remarks.}

We have here calculated branching-functions or coset characters
for a number of general coset models. We hope that we have
displayed how simple and straightforward the application of
the BRST invariant character is, to most general coset constructions.
Among those examples
we have, at least in part, verified results of other
authors, \cite{Bouwknegt-McCarthy-Pilch91}. They have under an assumption
obtained their branching functions. \\
\ind In the case of ref. \cite{Bouwknegt-McCarthy-Pilch91} the
assumption used is about the
cohomology of a Felder-like operator \cite{Felder89},\cite{Bernard-Felder90}
as well as the cohomology of the BRST-like operator.
In a later publication \cite{Bouwknegt-McCarthy-Pilch92:2} it is,
however, proven
that if the assumption of the Felder-like operator holds
then the cohomology of the BRST-like operator may be computed. It would
certainly be interesting to know if the remaining assumption is
completely validated due to the correctness of the character
or if there is still some freedom left. \\
\ind An obvious step in a more general understanding of coset models
would be to analyze non-compact groups. In order to do this
one would have to re-analyze the cohomology which does not
necessarily follow in a straightforward way from \cite{Hwang-Rhedin93}.
One of the problems is to choose unitary representations for
the group $G$. Those representations are not known for a general
non-compact group. What is also important is that for affine algebras
we must construct cosets that allow unitary representations, one
essentially must divide out the compact directions c.f. ref.
\cite{Dixon-Peskin-Lykken89}.
If we choose $G=SU(1,1)$, where the unitary
representations are known, the analysis should probably be
straightforward. \\
\ind Another open problem is constructions with arbitrary
rational levels. \\
\\
\\
\noindent{\bf{Acknowledgement}}

I am grateful to Stephen Hwang for suggesting this problem and for
discussions as well as explanations. I would also like to express
my gratitude to my wife, Anneli Rhedin, for typing the major
part of the manuscript.

\np

\np

\end{document}